\documentclass[a4paper,11pt]{article}

\usepackage{contribution}

\newcommand{\weblink}[2][]{%
    \ifthenelse{\equal{#1}{}}%
    {\textnormal{\url{#2}}}%
    {\textnormal{\href{#2}{#1}}}%
}

\newcommand{\acknowledgements}[1]{%
  \bigskip\bigskip
  \textsf{\textbf{\Large Acknowledgements}} \\[2ex]
  {#1}
  \bigskip
}

\def\beq{\begin{equation}}
\def\eeq#1{\label{#1}\end{equation}}
\def\eeqn{\end{equation}}

\def\beqa{\begin{eqnarray}}
\def\eeqa#1{\label{#1}\end{eqnarray}}
\def\eeqan{\end{eqnarray}}

\let\bar=\overbar

\def\Dslash{\not{\hbox{\kern-4pt $D$}}}
\def\dslash{\not{\hbox{\kern-2pt $\del$}}}

\def\msb{{\bar{\ssstyle M \kern -1pt S}}}

\newcommand{\contribution}[7][]{%
  \clearpage
  \thispagestyle{plain}
  \ifthenelse{\equal{#1}{}}
  {\hypersetup{pdftitle={#2}}}
  {\hypersetup{pdftitle={#1}}}
  \hypersetup{pdfauthor={{#3} {#4}}}
  {\centering\normalfont\LARGE\bfseries\sffamily #2 \par\nobreak}
  \lhead{}
  \chead{%
    \textit{\footnotesize XIV International Conference on Hadron Spectroscopy
      (\weblink[\textit{hadron2011}]{http://www.hadron2011.de}), 13-17 June 2011, Munich, Germany}%
  }
  \rhead{}
  \bigskip
  \begin{center}
    {#3} {#4}\ifthenelse{\equal{#6}{}}{}{\footnote{\weblink[#6]{mailto:#6}}}
    \ifthenelse{\equal{#7}{}}{}{#7} \\
    \textit{#5}
  \end{center}
  \bigskip
}

\renewcommand{\abstract}[1]{%
  \begin{center}
    \begin{minipage}{0.85\textwidth}
      \begin{footnotesize}
        #1
      \end{footnotesize}
    \end{minipage}
  \end{center}
  \bigskip
}

\begin{document}

%
%
%
%
%
{  

\newcommand{\chiral}{SU(3)$_R\, \times \, $SU(3)$_L$\,}
\newcommand{\im}{\text{Im }}
\newcommand{\Tr}{\text{Tr}}

%

\contribution[Magneto-Acoustic Coupling to Cyanide Reagents]  
{Meson-baryon interactions and baryon resonances}  
{Tetsuo}{Hyodo}  
{Department of Physics, Tokyo Institute of Technology, 
Meguro, Tokyo 152-8551, Japan}  
{hyodo@th.phys.titech.ac.jp}  
{}  
%

\abstract{%
  Meson-baryon interactions are the fundamental building blocks to study the structures of baryon resonances and meson properties in few-body nuclear systems. We review the recent progress in the investigation of the meson-baryon interaction in the strangeness $S=-1$ sector and the structure of the $\Lambda(1405)$ resonance. In particular, we present an attempt to construct a realistic $\bar{K}N$-$\pi\Sigma$ interaction in chiral SU(3) dynamics in response to the precise measurement of the kaonic hydrogen, and discuss the subthreshold extrapolation of the $\bar{K}N$ interaction with the information of the $\pi\Sigma$ channel.
}
%


\section{Introduction}

Hadrons are the asymptotic degrees of freedom of QCD at low energy due to color confinement, and the interactions among hadrons exhibit rich phenomena in the nonperturbative regime of the strong interaction. For instance, hadronic interactions in some sectors are so strong that two-body resonances are generated by hadronic dynamics~\cite{Dalitz:1959dn}. Such hadronic molecular states can also be found in the heavy quark sector, in which many new exotic mesons have been recently observed~\cite{Swanson:2006st}. The study of the hadron interactions and possible composite structures will elucidate a novel construction of hadrons and will open a new perspective in hadron spectroscopy. 

Hadronic interactions have been studied in various approaches. A traditional method for the hadron-hadron interaction is the meson-exchange picture~\cite{Machleidt:1987hj}. One of the recent achievements is the extraction of the inter-hadron forces from lattice QCD~\cite{Ishii:2006ec,Torok:2009dg}. For scattering systems including the Nambu-Goldstone bosons, on the other hand, chiral symmetry serves as a guiding principle to construct hadron interactions. A series of works based on chiral dynamics~\cite{Kaiser:1995eg,Oller:2000fj,Hyodo:2011ur} has demonstrated that the combination of the chiral low energy interaction with the unitarity condition of the scattering amplitude provides a unique tool to study the hadron interactions and resonances.

Here we consider meson-baryon scatterings with baryon resonances based on chiral SU(3) dynamics. Among others, we focus on the recent active discussions on the negative parity $\Lambda(1405)$ resonance with strangeness $S=-1$ and isospin $I=0$. There have been a long-standing debate on its structure; a three-quark state in the constituent quark model picture~\cite{Isgur:1978xj} vs. a $\bar{K}N$ bound state in the $\pi\Sigma$ continuum in the meson-baryon picture~\cite{Dalitz:1967fp}. When the $\Lambda(1405)$ resonance is regarded as a $\bar{K}N$ bound state, the interaction between $\bar{K}$ and $N$ is attractive enough to generate a bound state below the threshold. This picture motivates the study of the antikaon bound states in nuclei, the $\bar{K}$-nuclei~\cite{Akaishi:2002bg}. In this way, the study of the $\Lambda(1405)$ resonance has large impact on various fields of the strangeness sector of the nonperturbative QCD.
In section~\ref{sec:L1405} we introduce the formulation of chiral SU(3) dynamics, and discuss the structure of the $\Lambda(1405)$ resonance from various aspects. In section~\ref{sec:realistic}, we summarize the current status of the experimental investigations of the $\bar{K}N$-$\pi\Sigma$ amplitude and show possible future directions to be pursued for the construction of a realistic $\bar{K}N$-$\pi\Sigma$ interaction.

\section{$\Lambda(1405)$ in meson-baryon scattering}
\label{sec:L1405}

In this section we overview the framework of the chiral SU(3) dynamics for the meson-baryon scattering~\cite{Kaiser:1995eg,Oller:2000fj} in which the low energy chiral interaction is combined with the unitarity of the scattering amplitude. A detailed formulation of the model is given in Ref.~\cite{Hyodo:2011ur}. We then discuss the structure of the $\Lambda(1405)$ resonance in this approach, paying attention to the pole structure in the complex energy plane.

\subsection{Chiral SU(3) dynamics}
\label{subsec:ChU}

Regarding the light pseudoscalar mesons ($\pi$, $K$, and $\eta$) as the Nambu-Goldstone (NG) bosons associated with the spontaneous breaking of chiral \chiral symmetry, we can describe the low energy dynamics of the NG bosons in chiral perturbation theory~\cite{Gasser:1985gg}, which is an effective field theory based on the low energy expansion of the QCD Green's function. The chiral low energy theorems in current algebra are concisely encoded as the leading order results of the perturbative expansion. 

In chiral perturbation theory for meson-baryon systems, the covariant derivative of the baryon kinetic term provides the meson-baryon four-point vertex as
\begin{align*}
    \mathcal{L}^{\text{WT}}
    =\frac{1}{4f^2}\Tr \left(\bar{B}i\gamma^{\mu}[
    \Phi\partial_{\mu}\Phi
    -(\partial_{\mu}\Phi)\Phi
    ,B] \right) ,
\end{align*}
where $\Phi$ ($B$) is the octet meson (baryon) field and $f$ is the meson decay constant. From this interaction Lagrangian, we obtain the meson-baryon contact interaction called Weinberg-Tomozawa (WT) term, which takes on the following form after the $s$-wave projection:
\begin{equation}
    V_{i j}^{\text{WT}}(W) 
    = -  \frac{C_{i j}}{4 f^2}(2W - M_{i}-M_{j}) 
    \sqrt{\frac{M_i+E_i}{2M_i}}\sqrt{\frac{M_j+E_j}{2M_j}} ,
    \label{eq:WTtermswave}
\end{equation}
where $W$ is the total energy of the center-of-mass frame, $i$ labels the meson-baryon channel, and $M_{i}$ ($E_{i}$) is the mass (energy) of the baryon in channel $i$. The coupling strength $C_{ij}$ is determined by the flavor structure of the meson-baryon channel and the group theoretical argument~\cite{Hyodo:2006yk}. Equation~\eqref{eq:WTtermswave} provides the meson-baryon scattering length in accordance with the chiral low energy theorem~\cite{Weinberg:1966kf}.

In addition to the WT term, there are Born terms at $\mathcal{O}(p)$ and contact terms at $\mathcal{O}(p^{2})$~\cite{Borasoy:2004kk}, so the amplitude in chiral perturbation theory at next-to-leading order is given by
\begin{align}
    V(W)
    =& 
    V^{\text{WT}}(W)
    +V^{\text{Born}}(W)
    +V^{\text{NLO}}(W) .
    \nonumber
\end{align}
They are systematically calculated by the diagrams shown in Figure~\ref{fig:ChPT}. The Born terms mainly contribute to the $p$-wave amplitude and their $s$-wave component is in the higher order in the nonrelativistic expansion than the WT term. The next-to-leading-order (NLO) terms contain seven low energy constants which are not determined by the symmetry argument. Thus, for a qualitative discussion on the structure of the $\Lambda(1405)$ resonance, it is sufficient to adopt the WT interaction as the main component. In section~\ref{subsec:L1405}, we consider the structure of the $\Lambda(1405)$ resonance using the model with only the WT term. We will include the Born and the NLO terms in section~\ref{subsec:SIDDHARTA} for the quantitative discussion to construct a realistic $\bar{K}N$-$\pi\Sigma$ scattering amplitude.

\begin{figure}[tb]
  \begin{center}
    \includegraphics[width=0.8\textwidth]{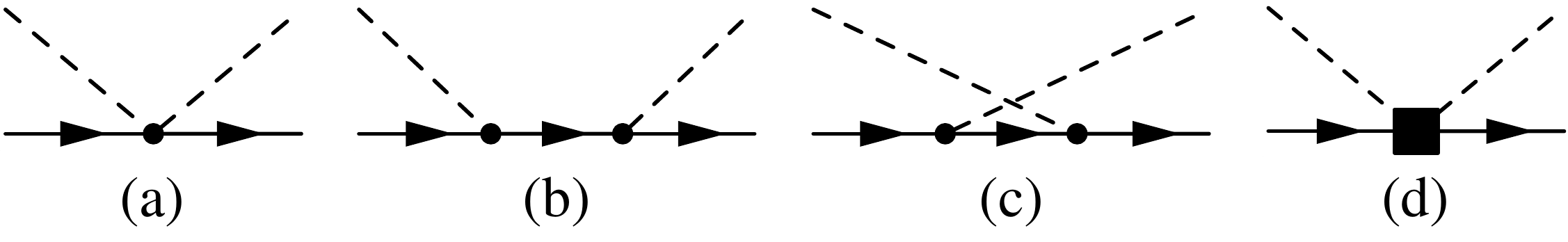}
    \caption{Feynman diagrams for the meson-baryon interactions in chiral perturbation theory. (a) Weinberg-Tomozawa interaction, (b) s-channel Born term, (c) u-channel Born term, and (d) Next-to-leading-order (NLO) interaction. The dots represent the $\mathcal{O}(p)$ vertices while the square denotes the $\mathcal{O}(p^2)$ vertex.}
     \label{fig:ChPT}
  \end{center}
\end{figure}

Chiral perturbation theory well describes the low energy meson-baryon scattering, but the unitarity condition is not always guaranteed because the amplitude is proportional to the meson momentum. To recover the unitarity, we make use of the dispersion relation. The unitarity condition leads to the optical theorem for the forward scattering amplitude $T(s)$:
\begin{equation}
    \im T^{-1}(s) = \frac{\rho(s)}{2}
    \quad
    \text{for } s\geq s_+
    \label{eq:Tinv} ,
\end{equation}
where $s=W^{2}$ is the Mandelstam variable and the two-body phase space factor is given by $\rho(s)=M\sqrt{(s-s_-)(s-s_+)}/(4\pi s)$ with $s_{\pm}=(M\pm m)^2$. Since the scattering amplitude is analytic in the complex energy plane except for possible poles (CDD poles), we can write the inverse scattering amplitude using dispersion relation as
\begin{equation}
    T^{-1}(W) = \sum_{n}\frac{R_n}{W-W_n}
    +\tilde{a}(s_0)
    +\frac{s-s_0}{2\pi}\int_{s_+}^{\infty}ds^{\prime}
    \frac{\rho(s^{\prime})}
    {(s^{\prime}-s)(s^{\prime}-s_0)} ,
    \nonumber
\end{equation}
which is consistent with the optical theorem~\eqref{eq:Tinv}. Single subtraction is performed at the subtraction point $s_0$ with the subtraction constant $\tilde{a}(s_0)$ to tame the divergence. $W_{n}$ and $R_{n}$ represent the position and the residue of the CDD poles, which cannot be determined within the scattering theory. Noting that the dispersion integral on the unitarity cut can be regarded as (the finite part of) the meson-baryon loop function $G(W)$, we can write the general form of the scattering amplitude as 
\begin{align}
    T(W) 
    =& 
    [\mathcal{T}^{-1}(W)
    -G(W)]^{-1} ,
    \label{eq:Tgen}
\end{align}
where $\mathcal{T}^{-1}(W)$ expresses the CDD pole contributions. The explicit form of the loop function is given by a diagonal matrix as
\begin{align*}
   G_k(W) 
   =
   &\frac{1}{(4\pi)^{2}} 
   \Bigl\{a_k(\mu)+\ln\frac{M_k^{2}}{\mu^{2}}
   +\frac{m_k^{2}-M_k^{2}+W^{2}}{2W^{2}}
   \ln\frac{m_k^{2}}{M_k^{2}}
   \nonumber\\
   &+\frac{\bar{q}_k}{W}
   [\ln(W^{2}-(M_k^{2}-m_k^{2})+2W\bar{q}_k)
   +\ln(W^{2}+(M_k^{2}-m_k^{2})+2W\bar{q}_k) 
   \nonumber\\
   &
   -\ln(-W^{2}+(M_k^{2}-m_k^{2})+2W\bar{q}_k)
   -\ln(-W^{2}-(M_k^{2}-m_k^{2})+2W\bar{q}_k)
   ]\Bigr\} ,  
\end{align*}
with $\bar{q}=\sqrt{(s-s_-)(s-s_+)}/(2\sqrt{s})$. $a_{k}(\mu)$ represents the subtraction constant which plays a role of the ultraviolet cutoff. To determine $\mathcal{T}(W)$, we match the loop expansion of the general form of the amplitude~\eqref{eq:Tgen} with that in chiral perturbation theory~\cite{Oller:2000fj}. Since the meson-baryon loop contribute as $\mathcal{O}(p^{3})$ in chiral perturbation theory, we can identify the $\mathcal{T}(W)$ function as the tree-level amplitude up to $\mathcal{O}(p^{2})$:
\begin{align}
    \mathcal{T}(W)
    =& 
    V^{\text{WT}}(W)
    +V^{\text{Born}}(W)
    +V^{\text{NLO}}(W) +\dots.
    \nonumber
\end{align}
where ellipsis denotes the higher order contributions. In this way, we can construct the scattering amplitude which is consistent with the unitarity condition and exhibits the correct low energy behavior required by chiral symmetry.

This form of the amplitude can be regarded as the solution of the scattering equation with the interaction kernel derived in chiral perturbation theory. In other words, we regard the meson-baryon tree-level amplitude as an interaction kernel of the scattering equation, and the meson-baryon amplitude is obtained by solving the scattering equation $T=V+VGT$ as illustrated in Figure~\ref{fig:LSE}. 

\begin{figure}[tb]
  \begin{center}
    \includegraphics[width=0.8\textwidth]{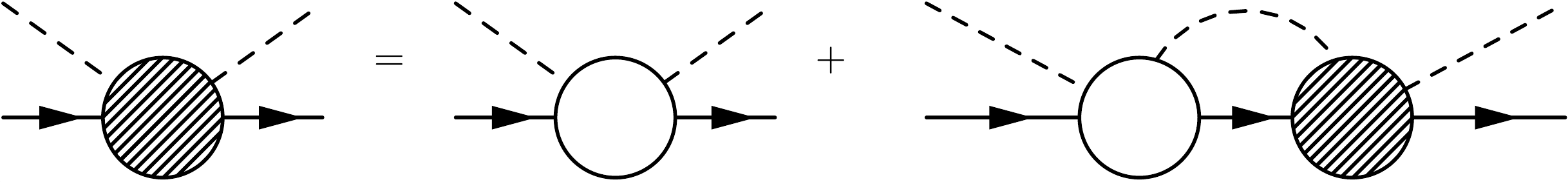}
    \caption{Schematic illustration of the scattering equation. The shaded (empty) blob represents the $T$-matrix (potential $V$). The loop function $G$ is expressed by the intermediate meson-baryon loop.}
     \label{fig:LSE}
  \end{center}
\end{figure}
%

\subsection{The structure of the $\Lambda(1405)$ resonance}
\label{subsec:L1405}

We have constructed the framework of chiral SU(3) dynamics. In this section, we apply this framework to the strangeness $S=-1$ and isospin $I=0$ meson-baryon scattering, and study the property of the $\Lambda(1405)$ resonance. In this sector, four channels ($\bar{K}N$, $\pi\Sigma$, $\eta\Lambda$, and $K\Xi$) participate in the coupled-channel scattering. For the clarity of the mechanism, we adopt the simplest model with the WT interaction. It is shown that this model successfully reproduces the observables in the $K^{-}p$ scattering and the mass spectrum of the $\Lambda(1405)$ resonance~\cite{Hyodo:2011ur}.

In the meson-baryon scattering amplitude, baryon resonances are expressed as pole singularities in the second Riemann sheet of the complex energy plane. Close to the resonance energy region, the scattering amplitude can be written as the Breit-Wigner term plus a slowly varying background term:
\begin{equation}
    T_{ij}(W) = \frac{g_ig_j}{W-M_R+i\Gamma_R/2} 
    + T_{ij}^{\text{BG}}(W) ,
    \nonumber
\end{equation}
where $M_{R}$ and $\Gamma_{R}$ are the mass and width of the resonance state, respectively, and $g_{i}$ stands for the coupling strength to the channel $i$. In this way, the properties of resonances can be read off from the pole structure of the amplitude.

In the study of $\Lambda(1405)$, it is found that there are two poles in the resonance energy region~\cite{Jido:2003cb}. This is illustrated in Figure~\ref{fig:pole1405} by plotting the absolute value of the scattering amplitude in the complex energy plane. We observe one bump structure on the real axis, which is influenced by two independent poles in the complex plane. By calculating the residues of the poles, it is shown that the higher energy pole strongly couples to the $\bar{K}N$ channel while the lower energy pole has larger coupling strength to the $\pi\Sigma$ channel. Because of the different coupling strengths, the scattering amplitude of $\bar{K}N\to\pi\Sigma$ and that of $\pi\Sigma\to\pi\Sigma$ are affected by two poles with different weights. As a consequence, the resonance shape of the $\Lambda(1405)$ resonance in the $\pi\Sigma$ mass spectrum may depend on the reaction process~\cite{Jido:2003cb}.

\begin{figure}[tb]
  \begin{center}
    \includegraphics[width=0.5\textwidth]{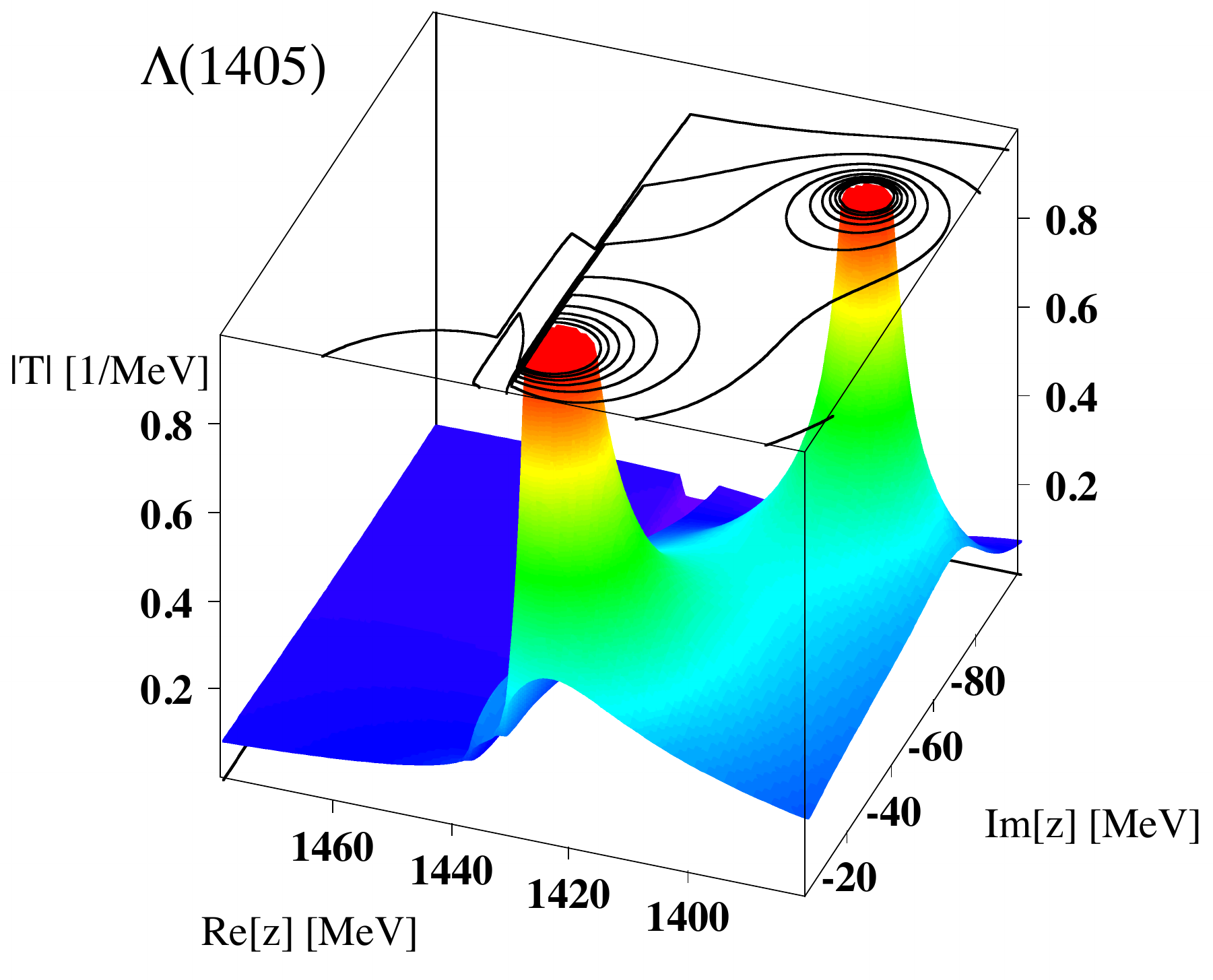}
    \caption{The absolute value of the scattering amplitude around the $\Lambda(1405)$ energy region in the complex plane.}
    \label{fig:pole1405}
  \end{center}
\end{figure}

The physical origin of the double-pole structure can be traced back to the two attractive components of the chiral interaction~\cite{Hyodo:2007jq}. The explicit form of the $C_{ij}$ coefficients of the WT interaction~\eqref{eq:WTtermswave} for $S=-1$ and $I=0$ channel is
\begin{equation}
    C_{ij}
    = 
    \begin{pmatrix}
    3 & -\sqrt{\dfrac{3}{2}} & \dfrac{3}{\sqrt{2}} & 0  \\
      & 4 & 0 & \sqrt{\dfrac{3}{2}} \\
      &   & 0 & -\dfrac{3}{\sqrt{2}} \\
      &   &   & 3 \\
    \end{pmatrix}
    \hspace{-1cm}
    \begin{matrix}
    \phantom{\sqrt{\dfrac{3}{2}}}& \bar{K}N \\ 
    \phantom{\sqrt{\dfrac{3}{2}}}& \pi\Sigma \\
    \phantom{\dfrac{3}{\sqrt{2}}}& \eta\Lambda \\
    \phantom{3} & K\Xi
    \end{matrix}
    .
    \nonumber
\end{equation}
A positive value corresponds to the attractive interaction, so we find from the diagonal components that both the $\bar{K}N$ and $\pi\Sigma$ channels are attractive. When we solve the scattering equation by eliminating the off-diagonal components, the $\bar{K}N$ channel develops one bound state below the threshold, and the $\pi\Sigma$ channel generates a resonance above the threshold. This is illustrated in Fig.~\ref{fig:polesingle} by plotting these pole positions together with those in the full coupled-channel model. It is clear from the figure that the higher energy pole originates in the $\bar{K}N$ bound state and the lower energy pole is evolved from the $\pi\Sigma$ resonance. Identifying the origin of the poles as a $\bar{K}N$ bound state and a $\pi\Sigma$ resonance is reasonable from the strong coupling of the higher (lower) energy pole to the $\bar{K}N$ ($\pi\Sigma$) channel. This analysis suggests that the $\Lambda(1405)$ resonance is realized as a Feshbach resonance in the resonating open channel~\cite{PRA}

\begin{figure}[tb]
  \begin{center}
    \includegraphics[width=0.5\textwidth]{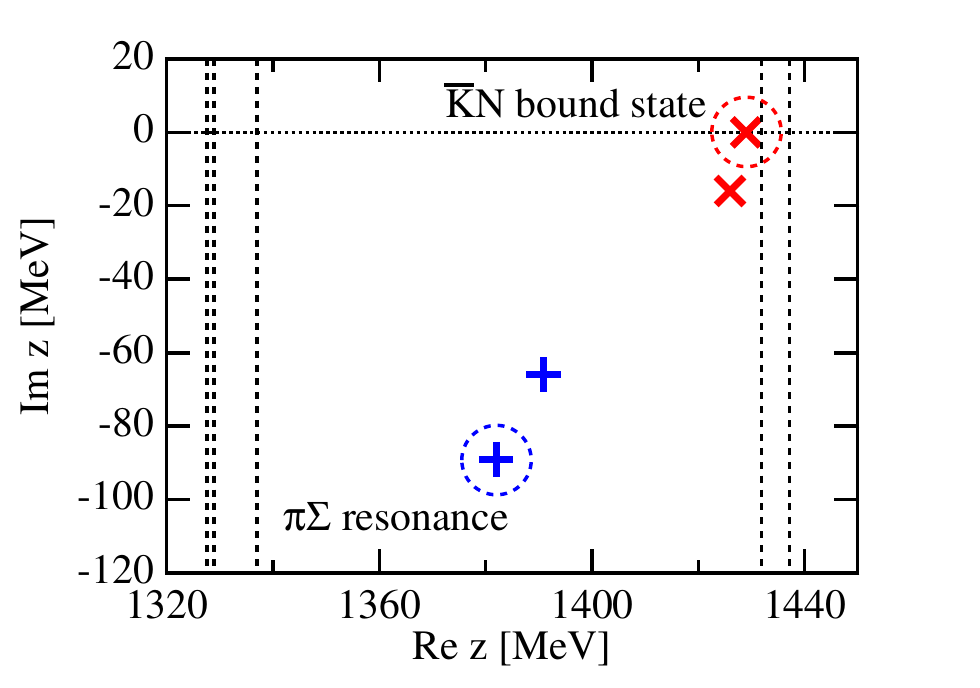}
    \caption{Pole positions of the scattering amplitude for $\Lambda(1405)$ in the complex energy plane. Crosses enclosed by circles represent poles obtained by switching off the transition couplings. Vertical dashed lines indicate the meson-baryon threshold energies.}
     \label{fig:polesingle}
  \end{center}
\end{figure}

Given the different nature of the coupling strengths, we notice that the pole structure of $\Lambda(1405)$ has an important phenomenological consequence in the $\bar{K}N$ interaction~\cite{Hyodo:2007jq}. We find that the $\bar{K}N$ interaction generates a quasi-bound state around 1420 MeV, rather than the nominal resonance position of 1405 MeV, and the physical $\Lambda(1405)$ is formed cooperatively with the attraction in the $\pi\Sigma$ channel. Thus, the strength of the $\bar{K}N$ interaction is closely related to the precise pole position(s) of the $\Lambda(1405)$ resonance. This issue will be further discussed in section~\ref{sec:realistic}, in relation with the quantitative refinements of the $\bar{K}N$-$\pi\Sigma$ interaction.

Before closing this section, let us mention several recent studies on the internal structure of the $\Lambda(1405)$ resonance. The origin of resonances is studied in Ref.~\cite{Hyodo:2008xr} through the renormalization procedure. As explained in section~\ref{subsec:ChU}, the CDD pole contribution should be included in the interaction kernel $V$, but certain choice of the subtraction constants in the loop function may introduce a seed of resonance in the loop function. To eliminate the CDD pole contribution from the loop function, the natural renormalization scheme is introduced (the meaning of this scheme is further discussed from the viewpoint of the compositeness~\cite{Compositeness}). Using the phenomenological fitting and the natural renormalization scheme, it is shown that the $\Lambda(1405)$ is dominated by the meson-baryon molecular structure. 

This picture is further confirmed from different aspects. The study of the $N_{c}$ scaling shows that the three-quark component in the $\Lambda(1405)$ resonance is small~\cite{Hyodo:2007np}. Through the evaluation of the electromagnetic properties~\cite{Sekihara:2008qk}, the spatial size of the $\Lambda(1405)$ is found to be much larger than the ground state hadrons which are presumably dominated by the three-quark structure. All these studies consistently indicate the dominance of the meson-baryon molecular structure of the $\Lambda(1405)$ resonance.

\section{Toward a realistic meson-baryon interaction}
\label{sec:realistic}

We have been discussing the structure of the $\Lambda(1405)$ resonance from the viewpoint of hadron spectroscopy. One of the central issues in the strangeness nuclear physics is the exploration of the kaonic nuclei. Since the $\Lambda(1405)$ resonance is located just below the $\bar{K}N$ threshold, its structure is closely related with the $\bar{K}N$ interaction~\cite{Hyodo:2007jq}. There are several theoretical studies on the $\bar{K}NN$ system~\cite{Shevchenko:2006xy}. The results of these studies show that the $\bar{K}NN$ three-body system will develop a quasi-bound state below the threshold with a large width. On the other hand, quantitative estimation of the binding energy does not converge with each other. 

To illustrate the current situation of the $\bar{K}N$ interaction, here we list the present experimental database which can be used to constrain the theoretical models:
\begin{itemize}
\item \textit{above} the $\bar{K}N$ threshold: total cross sections of $K^{-}p$ scattering into $K^{-}p$, $\bar{K}^{0}n$, $\pi^{0}\Sigma^{0}$, $\pi^{+}\Sigma^{-}$, $\pi^{-}\Sigma^{+}$, and $\pi^{0}\Lambda$ channels.

\item \textit{at} the $\bar{K}N$ threshold: threshold branching ratios and the $\bar{K}N$ scattering lengths.

\item \textit{below} the $\bar{K}N$ threshold: $\pi\Sigma$ invariant mass spectra.
\end{itemize}
At this point, it should be noted that the relevant energy region of the study of the bound antikaon in nuclei is below the $\bar{K}N$ threshold. Thus, the main reason of the discrepancy in the predictions of the $\bar{K}NN$ bound state can be attributed to the lack of the experimental information, especially in the region below the $\bar{K}N$ threshold. To increase the precision of the model calculation, we should accumulate 1) the $\bar{K}N$ data at the threshold and 2) the $\pi\Sigma$ data below the $\bar{K}N$ threshold, as we explain below.

The $\bar{K}N$ data at the threshold is the direct information of the $\bar{K}N$ channel with the lowest possible energy. At present, threshold branching ratios are measured with good accuracy~\cite{Tovee:1971ga}. In addition to that, the $K^{-}p$ scattering length is very important to constrain the $\bar{K}N$ interaction. Recent experimental efforts enable us to extract the precise value of this scattering length~\cite{Bazzi:2011zj}. 

Below the threshold, there is no direct information of the $\bar{K}N$ channel, and this region should be constrained indirectly through the data in the $\pi\Sigma$ channel. Many recent experiments report the $\pi\Sigma$ mass spectra for the $\Lambda(1405)$ energy region. Although they are important subthreshold informations of the $\bar{K}N$ interaction, usually the normalization of the spectrum is not known and several interference effects in experiments may modify the relevant structure~\cite{Hyodo:2011ur}. Instead, it is pointed out that the $\pi\Sigma$ threshold observables (scattering length and effective range) are also useful to constrain the subthreshold extrapolation of the $\bar{K}N$ interaction~\cite{Ikeda:2011dx}. The scattering length is given as a number at fixed energy, so it is unambiguously incorporated in the extrapolation procedure.

In this circumstance, it is of great relevance to construct a realistic $\bar{K}N$-$\pi\Sigma$ interaction using new experimental information, and to provide future directions of the study of meson-baryon interaction. We present a comprehensive analysis with the next-to-leading order chiral interaction in section~\ref{subsec:SIDDHARTA}, by use of the new precise measurement of the kaonic hydrogen. In section~\ref{subsec:piSigma}, we discuss the possibility of determining the $\pi\Sigma$ scattering lengths through the analysis of the hadronic decays of $\Lambda_{c}$.

\subsection{Improved constraints from new $\bar{K}N$ threshold data}
\label{subsec:SIDDHARTA}

Very recently, SIDDHARTA collaboration has reported a precise measurement of the 1s level of the kaonic hydrogen~\cite{Bazzi:2011zj}. The reported values of the energy shift $\Delta E$ and width $\Gamma$ are 
\begin{equation}
    \Delta E = 283\pm36(stat)\pm6(syst)\textrm{ eV} ,
    \quad \Gamma = 541\pm89(stat)\pm22(syst)\textrm{ eV} .
    \label{eq:SIDDHARTA}
\end{equation} 
The error bars are significantly reduced from previous measurements. These values are related to the (complex) $K^{-}p$ scattering length through the improved Deser-Trueman formula~\cite{Meissner:2004jr}. Thus, this new information can be used to constrain the real and imaginary parts of the $K^{-}p$ scattering amplitude at the $\bar{K}N$ threshold. 

A systematic $\chi^{2}$ analysis with these new constraints is performed in the framework of chiral SU(3) dynamics including NLO terms~\cite{IkedaWeise}. Three models with different interaction kernels are examined: WT model (with the Weinberg-Tomozawa term $V = V^{\text{WT}}$), WTB model (with the Weinberg-Tomozawa and Born terms $V= V^{\text{WT}}+V^{\text{Born}}$), and NLO model (all the terms $V = V^{\text{WT}}+V^{\text{Born}}+V^{\text{NLO}}$). The subtraction constants (and the low energy constants in the NLO model) are fitted to the new data of the kaonic hydrogen~\eqref{eq:SIDDHARTA}, the threshold branching ratios, and the $K^{-}p$ total cross sections. 

The results of the fitting are summarized in Table~\ref{tab:result}. Reasonable agreement with data is obtained in all cases, but in the WT and WTB models, some of the subtraction constants deviate from the natural value~\cite{Hyodo:2008xr}. The NLO model gives the best fit ($\chi^{2}$/d.o.f.$< 1$) and well reproduces the SIDDHARTA result~\eqref{eq:SIDDHARTA}, with natural-size subtraction constants~\cite{IkedaWeise}. Since the total cross section data is included in the $\chi^{2}$ analysis, the new result~\eqref{eq:SIDDHARTA} is shown to be consistent with the cross section data (see Refs.~\cite{Borasoy:2004kk} for the comparison with previous measurements of the kaonic hydrogen).
 
\begin{table}[tb]
  \begin{center}
    \begin{tabular}{l|cccc}  
      Model & WT & WTB & NLO & Experiment~\cite{Bazzi:2011zj} \\
      \hline
      $\Delta E$ [eV]  & 373 & 377 & 306 & $283\pm 42$ \\
      $\Gamma$ [eV]    & 495 & 514 & 591 & $541\pm 111$ \\
      \hline
      $\chi^{2}$/d.o.f & 1.12 & 1.15 & 0.96 &  \\
      \hline
      pole positions [MeV] & $1422-16i$ & $1421-17i$ & $1424-26i$ &  \\
      & $1384-90i$ & $1385-105i$ & $1381-81i$ &  \\
    \end{tabular}
    \caption{Results of the systematic $\chi^{2}$ analysis by chiral SU(3) dynamics for the $S=-1$ meson-baryon scattering~\cite{IkedaWeise}. Shown are the energy shift and width of the 1s state of the kaonic hydrogen ($\Delta E$ and $\Gamma$), $\chi^{2}$/d.o.f of the fitting, and the pole positions of the isospin $I=0$ amplitude in the $\bar{K}N$-$\pi\Sigma$ region.}
    \label{tab:result}
  \end{center}
\end{table}

The pole positions of the amplitude are shown in Table~\ref{tab:result}. All three models provide two poles in the energy region of the $\Lambda(1405)$ as discussed in section~\ref{subsec:L1405} and the double-pole structure of the $\Lambda(1405)$ resonance is confirmed. The uncertainty analysis, along the same line with Ref.~\cite{Borasoy:2006sr}, is also performed to check the stability of the fitting~\cite{IkedaWeise}.

\subsection{Information of $\pi\Sigma$ channel}
\label{subsec:piSigma}

As demonstrated in Ref.~\cite{Ikeda:2011dx}, the subthreshold extrapolation of the $\bar{K}N$ amplitude should not be discussed separately from the dynamics of the $\pi\Sigma$ channel. Although it is desirable to impose a constraint at the $\pi\Sigma$ threshold, so far no experimental information is available. 

Determination of the $\pi\Sigma$ scattering lengths in the $\Lambda_{c}$ decays is discussed in Ref.~\cite{Hyodo:2011js}, in the same strategy with the method developed for the $\pi\pi$ scattering length~\cite{Cabibbo:2004gq}. Because of the isospin breaking in the masses of $\pi$ and $\Sigma$, there is about 10 MeV mass difference in the transitions of $\pi^{+}\Sigma^- \to \pi^{-}\Sigma^+$, $\pi^{+}\Sigma^- \to \pi^{0}\Sigma^0$, and $\pi^{+}\Sigma^0 \to \pi^{0}\Sigma^+$. In the weak decays of the $\Lambda_{c}$ into $\pi\pi\Sigma$ channels, cusp structure appears at the energy of the former $\pi\Sigma$ channel in the spectrum of the latter $\pi\Sigma$ channel. This cusp structure reflects the transition amplitude between $\pi\Sigma$ channels at the threshold, which is nothing but the (off-diagonal) $\pi\Sigma$ scattering length. The value of the scattering length can be extracted from the expansion of the $\pi\Sigma$ spectrum around the threshold. It is shown that the substantial cusp effect should be observed in the spectrum, when the $\pi\Sigma$ interaction in $I=0$ is strongly attractive to generate a pole singularity around the threshold~\cite{Hyodo:2011js}.

This method can be applied to three different modes in the $\Lambda_{c}$ decays. Isospin decomposition of these channels reads
\begin{align}
    a^{-+}
    = & \frac{1}{3}a^{0}
    -\frac{1}{2}a^{1}
    +\frac{1}{6}a^{2}+\dots  ,
    \quad
    a^{00}
    =\frac{1}{3}a^{0}
    -\frac{1}{3}a^{2}+\dots  ,
    \quad
    a^{0+}
    =-\frac{1}{2}a^{1}
    +\frac{1}{2}a^{2}+\dots ,
    \nonumber 
\end{align}
where $a^{I}$ is the scattering length with isospin $I$ and ellipses represent the isospin breaking corrections. Since these three channels are not linearly independent as $a^{-+} - a^{00} = a^{0+} +\dots $, this method alone is not sufficient to determine all the isospin components, even if we extract the scattering lengths in all three channels. The last piece of the information may be completed by Lattice QCD~\cite{Torok:2009dg,IkedaLattice} which can determine the $\pi\Sigma$ scattering length in the $I=2$ channel.

\section{Summary}

We have studied the $\bar{K}N$-$\pi\Sigma$ interaction and the $\Lambda(1405)$ resonance based on chiral SU(3) symmetry and unitarity. It is shown that the chiral symmetry constrains the dynamics of the Nambu-Goldstone boson with hadrons at low energy, and the unitarity of the scattering amplitude enables us to construct the dynamical approach to the hadron scattering amplitude. We show that the $\Lambda(1405)$ resonance and the $S=-1$ meson-baryon scattering are well described in the framework of chiral SU(3) dynamics. The $\Lambda(1405)$ resonance exhibits a peculiar pole structure which is driven by the two attractive components of the chiral interaction in the $\bar{K}N$ and $\pi\Sigma$ channels.

The precise measurement of the kaonic hydrogen provides new constraints on the $\bar{K}N$-$\pi\Sigma$ amplitude. The chiral SU(3) dynamics at next-to-leading order is capable of accommodating this new information consistently with the total cross section data. The $\pi\Sigma$ scattering length can be an alternative observable to further constrain the $\bar{K}N$ interaction in the lower energy region. These activity will bring us a deep understanding of the $\Lambda(1405)$ resonance and will provide a baseline of the application of the $\bar{K}N$ interaction to the strangeness nuclear physics.

\acknowledgements{%
The author is grateful to Yoichi Ikeda, Daisuke Jido, Makoto Oka, and Wolfram Weise for many stimulating discussions and fruitful collaborations. He thanks the support from the Global Center of Excellence Program by MEXT, Japan through the Nanoscience and Quantum Physics Project of the Tokyo Institute of Technology. 
This work was partly supported by the Grant-in-Aid for Scientific Research from MEXT and JSPS (No.\ 21840026).
}


%

}  


\end{document}